\documentclass[sigconf]{acmart}
\usepackage{balance}
\def\BibTeX{{\rm B\kern-.05em{\sc i\kern-.025em b}\kern-.08emT\kern-.1667em\lower.7ex\hbox{E}\kern-.125emX}}
    
\copyrightyear{2019}
\acmYear{2019}
\setcopyright{acmcopyright}
\acmConference[KDD '19]{The 25th ACM SIGKDD Conference on Knowledge
Discovery and Data Mining}{August 4--8, 2019}{Anchorage, AK, USA}
\acmBooktitle{The 25th ACM SIGKDD Conference on Knowledge Discovery and Data
Mining (KDD '19), August 4--8, 2019, Anchorage, AK, USA}
\acmPrice{15.00}
\acmDOI{10.1145/3292500.3330668}
\acmISBN{978-1-4503-6201-6/19/08}

\usepackage{booktabs} 
\usepackage{graphicx}
\usepackage{courier}
\usepackage{makecell}
\usepackage{float}
\usepackage{bm}
\usepackage{multirow}
\usepackage[flushleft]{threeparttable}
\usepackage{flushend}
\usepackage{bbm}
\usepackage{mathrsfs}
\usepackage{enumitem}
\usepackage[ruled, linesnumbered]{algorithm2e} 
\newcommand{\ignore}[1]{}
\newcommand{\tabincell}[2]{\begin{tabular}{@{}#1@{}}#2\end{tabular}} 
\newcommand\ChangeRT[1]{\noalign{\hrule height #1}}
\newcommand{\ie}{\emph{i.e., }}

\newcommand{\eg}{\emph{e.g., }}
\newcommand{\etc}{\emph{etc}}

\usepackage{etoolbox}

\begin{document}
%
\title{Reinforcement Learning to Optimize Long-term User Engagement in Recommender Systems}

%

\author{Lixin Zou}
\authornote{Work performed during an internship at JD.com.}
\affiliation{\institution{Tsinghua University}}
\email{zoulx15@mails.tsinghua.edu.cn}

\author{Long Xia}
\affiliation{\institution{Data Science Lab, JD.com}}
\email{xialong@jd.com}

\author{Zhuoye Ding}
\affiliation{\institution{Data Science Lab, JD.com}}
\email{dingzhuoye@jd.com}

\author{Jiaxing Song}
\affiliation{\institution{Tsinghua University}}
\email{jxsong@mail.tsinghua.edu.cn}

\author{Weidong Liu}
\affiliation{\institution{Tsinghua University}}
\email{liuwd@mail.tsinghua.edu.cn}

\author{Dawei Yin}
\affiliation{\institution{Data Science Lab, JD.com}}
\email{yindawei@acm.org}

%
\renewcommand{\shortauthors}{Lixin et al.}

%
\begin{abstract}
Recommender systems play a crucial role in our daily lives. Feed streaming mechanism has been widely used in the recommender system, especially on the mobile Apps. The feed streaming setting provides users the interactive manner of recommendation in never-ending feeds. In such a manner, a good recommender system should pay more attention to user stickiness, which is far beyond classical instant metrics and typically measured by {\bf long-term user engagement}. Directly optimizing long-term user engagement is a non-trivial problem, as the learning target is usually not available for conventional supervised learning methods. Though reinforcement learning~(RL) naturally fits the problem of maximizing the long term rewards, applying RL to optimize long-term user engagement is still facing challenges: user behaviors are versatile to model, which typically consists of both instant feedback~(\eg clicks) and delayed feedback~(\eg dwell time, revisit); in addition, performing effective off-policy learning is still immature, especially when combining bootstrapping and function approximation.

To address these issues, in this work, we introduce a RL framework --- FeedRec to optimize the long-term user engagement. FeedRec includes two components: 1)~a Q-Network which designed in hierarchical LSTM takes charge of modeling complex user behaviors, and 2)~a S-Network, which simulates the environment, assists the Q-Network and voids the instability of convergence in policy learning. Extensive experiments on synthetic data and a real-world large scale data show that FeedRec effectively optimizes the long-term user engagement and outperforms state-of-the-arts.

\end{abstract}

%
%


%
\maketitle
\section{Introduction}
Recommender systems assist user in information-seeking tasks by suggesting goods~(\eg products, news, services) that best match users' needs and preferences. In recent feed streaming scenarios, users are able to constantly browse items generated by the never-ending feeds, such as the news streams in Yahoo News\footnote{https://ca.news.yahoo.com/}, the social streams in Facebook\footnote{https://www.facebook.com/}, and the product streams in Amazon\footnote{https://www.amazon.com/}. Specifically, interacting with the product streams, the users could click on the items and view the details of the items. Meanwhile, (s)he could also skip unattractive items and scroll down, and even leave the recommender system due to the appearance of many redundant or uninterested items. Under such circumstance, optimizing clicks will not be the only golden rule anymore.
It is critical to maximizing the users' satisfaction of interactions with the feed streams, which falls in two folds: instant engagement, \eg click, purchase; long-term engagement, say stickiness, typically representing users' desire to stay with the streams longer and open the streams repeatedly~\cite{lalmas:userengagement}.

However, most traditional recommender systems only focus on optimizing instant metrics (\eg click through rate \cite{Li:NARM},  conversion rate\cite{pradel:case}).
Moving more deeply with interaction, a good feed streaming recommender system should be able to not only bring about higher click through rate but also keep users actively interacting with the system, which typically is measured by long-term delayed metrics.
Delayed metrics usually are more complicated, including dwell time on the Apps, depth of the page-viewing, the internal time between two visits, and so on.
Unfortunately, due to the difficulty of modeling delayed metrics, directly optimizing the delayed metrics is very challenging.
While only a few preliminary work\cite{Wu:return} starts investigating the optimization of some long-term/delayed metrics, a systematical solution to optimize the overall engagement metrics is wanted.

Intuitively, reinforcement learning (RL), which was born to maximize long-term rewards, could be a unified framework to optimize the instant and long-term user engagement. 
Applying RL to optimize long-term user engagement itself is a non-trivial problem. As mentioned, the long-term user engagement is very complicated~(\ie measured in versatile behaviors, \eg dwell time, revisit), and would require a very large number of environment interactions to model such long term behaviors and build a recommendation agent effectively. As a result, building a recommender agent from scratch through real online systems would be prohibitively expensive, since numerous interactions with immature recommendation agent will harm user experiences, even annoy the users.
An alternative is to build a recommender agent offline through making use of the logged data, where the off-policy learning methods can mitigate the cost of the trial-and-error search. Unfortunately, current methods including Monte Carlo~(MC) and temporal-difference~(TD) have limitations for offline policy learning in realistic recommender systems: MC-based methods suffer from the problem of high variance, especially when facing enormous action space~(\eg billions of candidate items) in real-world applications; TD-based methods improve the efficiency by using bootstrapping techniques in estimation, which, however, is confronted with another notorious problem called {\it Deadly Triad}~(\ie the problem of instability and divergence arises whenever combining function approximation, bootstrapping, and offline training~\cite{Sutton:RL}). Unfortunately, state-of-the-art methods~\cite{zhao2018Pairwise,zhao:deepDPG} in recommender systems, which are designed with neural architectures, will encounter inevitably the {\it Deadly Triad} problem in offline policy learning.

To overcome the aforementioned issues of complex behaviors and offline policy learning, we here propose an RL-based framework, named FeedRec, to improve long-term user engagement in recommender systems.
Specifically, we formalize the feed streaming recommendation as a {\it Markov decision process} (MDP), and design a Q-Network to directly optimize the metrics of user engagement. To avoid the problem of instability of convergence in offline Q-Learning, we further introduce a S-Network, which simulates the environments, to assist the policy learning. 
In Q-Network, to capture the information of versatile user long term behaviors, a fine user behavior chain is modeled by LSTM, which consists of all rough behaviors, \eg click, skip, browse, ordering, dwell, revisit, \etc. When modeling such fine-grained user behaviors, two problems emerges:  the numbers for specific user actions is extremely imbalanced (\ie clicks is much fewer than skips)\cite{zhou2018micro}; and long-term user behavior is more complicated to represent. We hence further integrated hierarchical LSTM with temporal cell into Q-Network to characterize fine-grained user behaviors.

On the other hand, 
in order to make effective use of the historical logged data and avoid the {\it Deadly Triad} problem in offline Q-Learning, we introduce an environment model, called S-network, to simulate the environment and generate simulated user experiences, assisting offline policy learning.
We conduct extensive experiments on both the synthetic dataset and a real-world E-commerce dataset. The experimental results show the effectiveness of the proposed algorithm over the state-of-the-art baselines for optimizing long user engagement.

Contributions can be summarized as follow:
\begin{enumerate}[leftmargin=*]
\item We propose a reinforcement learning model --- FeedRec to directly optimize the user engagement (both instant and long term user engagement) in feed streaming recommendation.
\item To model versatile user behaviors, which typically includes both instant engagement (\eg click and order) and long term engagement (\eg dwell time, revisit, \etc), Q-Network with hierarchical LSTM architecture is presented.
\item To ensure convergence in off-policy learning, an effective and safe training framework is designed.
\item The experimental results show that our proposed algorithms outperform the state-of-the-art baseline. 
\end{enumerate}
\section{Related Work}\label{related work}
\subsection{Traditional recommender system} Most of the existing recommender systems try to balance the instant metrics and factors, \ie the diversity, the novelty in recommendations. From the perspective of the instant metrics, there are numerous works focusing on improving the users' implicit feedback clicks~\cite{Hidasi:sessionRNN,Li:NARM,wang2018path}, explicit ratings~\cite{Mnih:PMF,Rendle:FM,chang2017streaming}, and dwell time on recommended items~\cite{Yi:BeyondClicks}. In fact, the instant metrics have been criticized to be insufficient to measure and represent real engagement of users. As the supplementary, methods~\cite{Adomavicius:AggregateDIV,Ashkan:GreedyDIV,Cheng:AccurateDiversity} intended to enhance user's satisfaction through recommending diverse items have been proposed. However, all of these works can not model the iterative interactions with users. Furthermore, none of these works could directly optimize delayed metrics of long-term user engagement. 
\subsection{Reinforcement learning based recommender system} Contextual bandit solutions are proposed to model the interaction with users and handle the notorious explore/exploit dilemma in online recommendation~\cite{Li:contextualBandit,Wang:Factorization,Qin:contextual,Zeng:Online_context,he2019offpolicy}. On one hand, these contextual bandit settings assume that the user's interests remain the same or smoothly drift which can not hold under the feed streaming mechanism. On the other hand, although Wu et al.~\cite{Wu:return} proposed to optimize the delayed revisiting time, there is no systematical solution to optimizing delayed metrics for user engagement. Apart from contextual bandits, a series of MDP based models~\cite{Shani:MDP,Mahmood:Improving,Dulac:RL_large_action,Lu:partial,zou2019reinforcement,zhao2017list,zhao2018survey} are proposed in recommendation task. Arnold et al.~\cite{Dulac:RL_large_action} proposed a modified DDPG model to deal with the problem of large discrete action spaces. Recently, Zhao et al. combined pagewise, pairwise ranking technologies with reinforcement learning\cite{zhao2018Pairwise,zhao:deepDPG}. Since only the instant metrics are considered, the above methods fail to optimize delayed metrics of user engagement. In this paper, we proposed a systematically MDP-based solution to track user's interests shift and directly optimize both instant metrics and delayed metrics of user engagement.

\section{Problem Formulation}

\subsection{Feed Streaming Recommendation}
In the feed streaming recommendation, the recommender system interacts with a user $u\in \mathcal{U}$ at discrete time steps. At each time step $t$, the agent feeds an item $i_t$ and receives a feedback $f_t$ from the user, where $i_t \in \mathcal{I}$ is from the recommendable item set and $f_{t}\in \mathcal{F}$ is user's feedback/bevahior on $i_{t}$, including clicking, purchasing, or skipping, leaving, \etc. The interaction process forms a sequence $X_t=\{u,(i_{1},f_{1},d_{1}),\dots,(i_{t},f_{t},d_{t})\}$ with $d_t$ as the dwell time on the recommendation, which indicates user's preferences on the recommendation. Given $X_t$, the agent needs to generate the $i_{t+1}$ for next-time step with the goal of maximizing long term user engagement, \eg the total clicks or browsing depth. In this work, we focus on how to improving the expected quality of all items in feed streaming scenario.

\subsection{MDP Formulation of Feed Streams}
A MDP is defined by $M = \langle S , A , P , R , \gamma \rangle$, where $S$ is the state space, $A$ is the action space, $P : S \times A \times S \rightarrow \mathbb{R}$ is the transition function, $R : S \times A \rightarrow \mathbb { R }$ is the mean reward function with $r(s,a)$ being the immediate goodness of $(s,a)$, and $\gamma \in [ 0,1 ]$ is the discount factor. A (stationary) policy $\pi : S \times A \rightarrow [ 0,1 ]$ assigns each state $s\in S$ a distribution over actions, where $a \in A$ has probability $\pi (a|s)$. In feed streaming recommendation, $\langle S , A , P \rangle$ are set as follow:
\begin{itemize}[leftmargin=*]
	\item {\bf State $S$} is a set of states. We design the state at time step $t$ as the browsing sequence $s_t=X_{t-1}$. At the beginning, $s_1 = \{u\}$ just contains user's information. At time step $t$, $s_{t} = s_{t-1}\oplus \{(i_{t-1},f_{t-1},d_{t-1})\}$ is updated with the old state $s_{t-1}$ concentrated with the tuple of recommended item, feedback and dwell time $({i}_{t-1},{f}_{t-1},d_{t-1})$.
	\item {\bf Action $A$} is a finite set of actions. The actions available depends on the state $s$, denoted as $A(s)$. The $A(s_1)$ is initialized with all recalled items. $A(s_t)$ is updated by removing recommended items from $A(s_{t-1})$ and action $a_{t}$ is the recommending item $i_t$.
	\item {\bf Transition $P$}  is the transition function with $p \left( s_{t+1} | s_t , i_t \right)$ being the probability of seeing state $s_{t+1}$ after taking action $i_t$ at $s_t$. In our case, the uncertainty comes from user's feedback $f_t$ \emph{w.r.t.} $i_t$ and $s_t$. 
\end{itemize}

\subsection{User Engagement and Reward Function} 
As aforementioned, unlike traditional recommendation, instant metrics (click, purchase, \etc) are not the only measurements of the user engagement/satisfactory, and long term engagement is even more important, which is often measured in delayed metrics, \eg browsing depth, user revisits and dwells time on the system. Reinforcement learning provides a way to directly optimize both instant and delayed metrics through the designs of reward functions.

The reward function $R:S\times A\rightarrow \mathbb{R}$ can be designed in different forms. We here instantiate it linearly by assuming that user engagement reward $r_t(\bm{m}_t)$ at each step $t$ is in the form of weighted sum of different metrics:
\begin{equation}\label{eq:overall_reward}
	r_t = \bm{\omega}^{\top}\bm{m}_t,
\end{equation}
where $\bm{m}_t$ is a column vector consisted of different metrics, $\bm{\omega}$ is the weight vector. 
Next, we give some instantiations of reward function \emph{w.r.t.} both instant metrics and delayed metrics.

\paragraph{\textbf{Instant metrics}} In the instant user engagement, we can have clicks, purchase (in e-commerce), \etc. The shared characteristics of instant metrics are that these metrics are triggered instantly by the current action. We here take click as an example, the number of clicks in $t$-th feedback is defined as the metric for click $m^c_t$,
{\small
\[
	m^{c}_t	= \#clicks(f_t).
\]}

\paragraph{\textbf{Delayed metrics}}
The delayed metrics include browsing depth, dwell time on the system, user revisit, \etc. Such metrics are usually adopted for measuring long-term user engagement. The delayed metrics are triggered by previous behaviors, some of which even hold long-term dependency. We here provide two example reward functions for delayed metrics:
\paragraph{Depth metric.}
The depth of browsing is a special indicator that the feed streaming scenario differs from other types of recommendation due to the infinite scroll mechanism.
After viewing the $t$-th feed, the system should reward this feed if the user remained in the system and scrolled down. Intuitively, the metric of depth $m_t^d$ can be defined as:
\[
    m_t^d    = \#scans(f_t)
\]
where $\#scans(f_t)$ is the number of scans in the $t$-th feedback.

\paragraph{Return time metric.}
The user will use the system more often when (s)he is satisfied with the recommended items.
Thus, the interval time between two visits can reflect the user's satisfaction with the system.
The return time $m^r_t$ can be designed as the reciprocal of time:
\[
	m^{r}_t = \frac{\beta}{v^{r}},
\]
where $v^{r}$ represents the time between two visits and $\beta$ is the hyper-parameter.

From the above examples---click metric, depth metric and return time metric, we can clearly see $\bm{m}_t = [m^{c}_t, m^{d}_t, m^{r}_t]^\top$. Note that in MDP setting, cumulative rewards will be maximized, that is, we are actually optimizing total browsing depth, and frequency of visiting in the future, which typically are long term user engagement. 

\section{Policy Learning for Recommender Systems}
To estimate the future reward (\ie the future user stickiness), the expected long-term user engagement for recommendation $i_t$ is presented with the Q-value as,
{\small
\begin{eqnarray}
Q^\pi(s_t,i_t) =  \mathbb{E}_{i_{k}\sim\pi}[\underbrace{r_t}_{\text{current rewards}} + \underbrace{\sum_{k=1}^{T-t}\gamma^k r_{t+k}}_{\text{future rewards}}],
\end{eqnarray}}
where $\gamma$ is the discount factor to balance the importance of the current rewards and future rewards. The optimal $Q^\ast(s_t,i_t)$, having the maximum expected reward achievable by the optimal policy, should follow the optimal Bellman equation~\cite{Sutton:RL} as,
{\small
\begin{eqnarray}
Q ^ { * } ( s_t , i_t ) = \mathbb { E } _ {s_{t+1}} \left[ r_t + \gamma \max _ { i ^ { \prime } } Q ^ { * } \left( s_{t+1} , i^{\prime} \right) | s_{t} , i_{t} \right].
\end{eqnarray}}
Given the $Q ^ { * }$, the recommendation $i_t$ is chosen with the maximum $Q ^ { * } ( s_t , i_t )$. Nevertheless, in real-world recommender systems, with enormous users and items, estimating the action-value function $Q^\ast(s_t,i_t)$ for each state-action pairs is infeasible. Hence, it is more flexible and practical to use function approximation, \eg neural networks, to estimate the action-value function, \ie $Q^\ast(s_t,i_t)\approx Q(s_t,i_t;\theta_q)$. In practice, neural networks are excellent to track user's interests in recommendation~\cite{Hidasi:sessionRNN,Zheng:NACF,Li:NARM}. In this paper, we refer to a neural network function approximator with parameter $\theta_q$ as a Q-Network. The Q-Network can be trained by minimizing the mean-squared loss function, defined as follows:
{\small
\begin{eqnarray}\label{equ:dqn}
\ell(\theta_q) &=& \mathbb{E}_{(s_t,i_t,r_t,s_{t+1})\sim \mathcal{M}}\left[(y_t-Q(s_t,i_t;\theta_q))^2\right]\\ \nonumber
    y_t &=& r_t + \gamma \max_{i_{t+1}\in \mathcal{I}} Q(s_{t+1},i_{t+1};\theta_q),
\end{eqnarray}}
where $\mathcal{M}=\{(s_t,i_t,r_t,s_{t+1})\}$ is a large replay buffer storing the past feeds, from which samples are taken in mini-batch training. By differentiating the loss function with respect to $\theta_q$, we arrive at the following gradient:
{\small
\begin{eqnarray}\label{equ:qlearning}
\nonumber \nabla _ { \theta_q } \ell \left( \theta_q  \right) &=& \mathbb { E } _ { \left( s_t , i_t , r_t , s_{t+1} \right) \sim \mathcal {M} } \left[ ( r + \gamma \max _ { i_{t+1}  } Q  \left( s_{t+1} , i_{t+1} ; \theta_q \right) \right. \\
&& \left.\left. - Q \left( s_t , i_t;\theta_q\right) \right)\nabla _ { \theta_q  } Q \left( s_t , i_t ; \theta_q \right)\right] 
\end{eqnarray}}

In practice, it is often computationally efficient to optimize the loss function by stochastic gradient descent, rather than computing the full expectations in the above gradient.

\begin{figure}[!t]
    \centering
    \includegraphics[width=8.0cm]{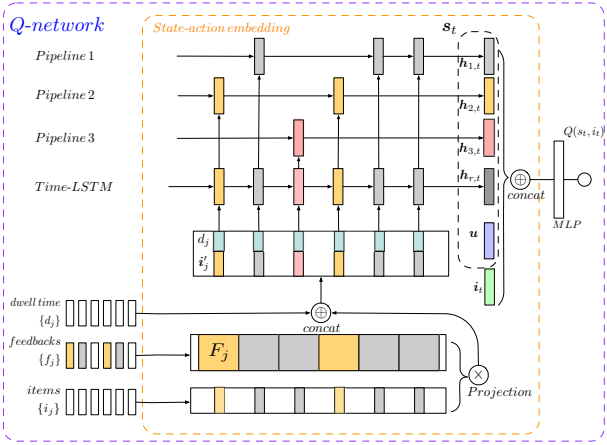}
    \caption{\small The architecture of Q-Network.}
    \label{fig:Q-Network}
\end{figure}

\subsection{The Q-Network}
The design of Q-Network is critical to the performances. In long term user engagement optimization, the user interactive behaviors is versatile (\eg not only click but also dwell time, revisit, skip, \etc), which makes modeling non-trivial. To effective optimize such engagement, we have to first harvest previous information from such behaviors into Q-Network.

\subsubsection{Raw Behavior Embedding Layer}
The purpose of this layer is to take all raw behavior information, related to long term engagement, to distill users' state for further optimization.
Given the observation $s_t=\{u,(i_1,f_1,d_1)\dots,(i_{t-1},f_{t-1},d_{t-1})\}$, we let $f_{t}$ be all possible types of user behaviors on $i_{t}$, including clicking, purchasing, or skipping, leaving \etc, while $d_t$ for the dwell time of the behavior. The entire set of $\{i_t\}$ are first converted into embedding vectors $\{\bm{i}_{t}\}$. To represent the feedback information into the item embedding, we project $\{\bm{i}_t\}$ into a feedback-dependent space by multiplying the embedding with a projection matrix as follow:
{\small
\begin{eqnarray}
    \bm{i}^{\prime}_{t} = F_{f_t}\bm{i}_{t}, \nonumber
\end{eqnarray}}
where $F_{f_t}\in \mathbb{R}^{H\times H}$ is a projection matrix for a specific feedback $f_{t}$. 
To futher model time information, in our work, a time-LSTM\cite{zhu2017next} is used to track the user state over time as:
{\small
\begin{eqnarray}\label{equ:time_lstm}
    \bm{h}_{r,t} = Time\text{-}LSTM(\bm{i}^{\prime}_{t},d_t),
\end{eqnarray}}
where Time-LSTM models the dwell time by inducing a time gate controlled by $d_t$ as follow:
{\small
\begin{eqnarray}
    \nonumber \bm{g} _ { t } &=& \sigma \left( \bm{i}^{\prime}_{t} W _ { i g } + \sigma  \left( d _ { t } W _ {gg} \right) + b _ { g } \right) \\
   \nonumber \bm{c} _ { t } &= & \bm{p} _ { t } \odot \bm{c} _ { t - 1 } + \bm{e} _ { t } \odot \bm{g} _ { t } \odot \sigma \left( \bm{i}^{\prime}_{t} W _ { i c } + \bm{h} _ { t - 1 } W _ { h c } + b _ { c } \right) \\ 
   \nonumber  \bm{o} _ { t } &= & \sigma ( \bm{i}^{\prime}_{t} W _ { i o } + d_ { t } W _ { d o } + \bm{h} _ { t - 1 } W _ { h o } + \bm{w} _ { c o } \odot \bm{c} _ { t } + b _ { o } ),
\end{eqnarray}}
where $\bm{c}_{t}$ is the memory cell. $\bm{g}_ { t }$ is the time dependent gate influencing the memory cell and output gate. $\bm{p} _ { t }$ is the forget gate. $\bm{e}_t$ is the input gate. $\bm{o} _ { t }$ is the output gate. $W_{\ast}$ and $b_{\ast}$ are the weight and bias term. $\odot$ is the element-wise product, $\sigma$ is the sigmoid function. Given the $\bm{c}_t$ and $\bm{o}_t$, the hidden state $\bm{h}_{r,t}$ is modeled as 
{\small
\begin{eqnarray}
    \nonumber \bm{h}_{r,t} = o_t\odot \sigma(c_t).
\end{eqnarray}}

\begin{figure}[!t]
    \centering
    \includegraphics[width=7cm]{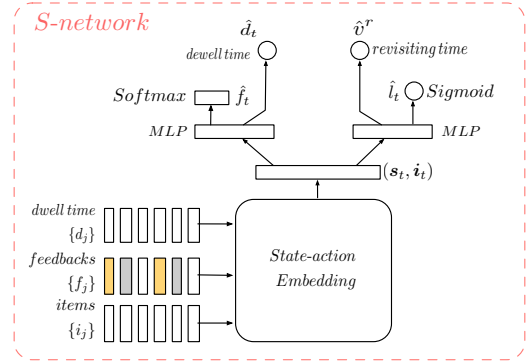}
    \caption{\small The architecture of S-Network.}
    \label{fig:s_network}
\end{figure}

\subsubsection{Hierarchical Behavior Layer}
To capture the information of versatile user behaviors, all rough behaviors are sequentially fed into raw Behavior Embedding Layer indiscriminate. In realistic, the numbers for specific user actions is extremely imbalanced (\eg clicks are fewer than skips). As a result, directly utilizing the output of raw Behavior Embedding Layer will cause the Q-Network losing the information from the sparse user behaviors, \eg purchase information will be buried by skips information. Moreover, each type of user behaviors has its own characteristics: click on an item usually represents the users' current preferences, purchase on an item may imply the shifting of user interest, and causality of skipping is a little complex, which could be casual browsing, neutral, or annoyed, \etc.

To better represent the user state, as shown in Figure \ref{fig:Q-Network}, we propose a hierarchical behavior layer added to the raw behaviors embedding layers, that the major user behaviors, such as click, skip, purchase are tracked separately with different LSTM pipelines as
{\small
\begin{eqnarray}
    \nonumber \bm{h}_{k,t} &=& \text{LSTM-k} (\bm{h}_{r,t}) \text{ if }f_t\text{ is the k-th behavior,}
\end{eqnarray}}
where different user's behaviors (\eg the k-th behavior) is captured by the corresponding LSTM layer to avoid intensive behavior dominance and capture specific characteristics. Finally, the state-action embedding is formed by concatenating different user's behavior layer and user profile as:
{\small
\begin{eqnarray}
    \nonumber \bm{s_t} = \text{concat}[\bm{h}_{r,t},\bm{h}_{1,t},\bm{h}_{\cdot,t},\bm{h}_{k,t},\bm{u}],
\end{eqnarray}}
where $\bm{u}$ is the embedding vector for a specific user.

\subsubsection{Q-value Layer}
The approximation of Q-value is accomplished by MLP with the input of  the dense state embedding and the item embedding as follow:
{\small
\begin{eqnarray}
    \nonumber Q(s_t,i_t;\theta_q) = MLP(\bm{s}_t, \bm{i}_t).
\end{eqnarray}}
The value of $\theta_q$ is updated by SGD with gradient calculated as Equation~(\ref{equ:qlearning}).

\subsection{Off-Policy Learning Task}
With the proposed Q-Learning based framework, we can train the parameters in the model through trial and error search before learning a stable recommendation policy. However, due to the cost and risk of deploying unsatisfactory policies, it is nearly impossible for training the policy online. An alternative way is to train a reasonable policy using the logged data $\mathcal{D}$, collecting by a logging policy $\pi_b$, before deploying. Unfortunately, the Q-Learning framework in Equation (\ref{equ:dqn}) suffers from the problem of {\it Deadly Trial}\cite{Sutton:RL}, the problem of instability and divergence arises whenever combining function approximation, bootstrapping and offline training.

To avoid the problem of instability and divergence in offline Q-Learning, we further introduce a user simulator (refers to as S-Network), which simulates the environment and assists the policy learning. Specifically, in each round of recommendation, aligning with real user feedback, the S-Network need to generate user's response $f_t$, the dwell time $d_t$, the revisited time $v^r$, and a binary variable $l_t$, which indicates whether the user leaves the platform. As shown in Figure \ref{fig:s_network}, the generation of simulated user's feedback is accomplished using the S-Network $S(\theta_s)$, which is a multi-head neural network. State-action embedding is designed in the same architecture as it in Q-Network, but has separate parameters. The layer~$(\bm{s}_t,\bm{i}_t)$ are shared across all tasks, while the other layers~(above $(\bm{s}_t,\bm{i}_t)$ in Figure~\ref{fig:s_network}) are task-specific. As dwell time and user's feedback are inner-session behaviors, the prediction of $\hat{\bm{f}}_t$ and $\hat{d}_t$ is calculated as follow,
{\small
\begin{eqnarray}
    \nonumber\hat{\bm{f}}_t &=& \text{Softmax}(W_f \bm{x}_f + b_f)\\
    \nonumber\hat{d}_t &=& W_d \bm{x}_f + b_d \\
    \nonumber\bm{x}_{f} &=& tanh (W_{xf} [\bm{s}_t, \bm{i}_t] + \bm{b}_{xf})
\end{eqnarray}}
where $W_\ast$ and $b_\ast$ are the weight and bias term. $[\bm{s}_t, \bm{i}_t]$ is the concentration of state action feature.
The generation of revisiting time and leaving the platform (inter-session behaviors) are accomplished as
{\small
\begin{eqnarray}
    \nonumber\hat{\bm{l}}_t &=& \text{Sigmoid}(\bm{x}_f^\top \bm{w}_l+ b_l)\\
    \nonumber\hat{v}^r &=& W_v \bm{x}_l + b_d \\
    \nonumber\bm{x}_{l} &=& tanh (W_{xl} [\bm{s}_t, \bm{i}_t] + \bm{b}_{xl}).
\end{eqnarray}}

{\footnotesize
\begin{algorithm}[!hpt]
\SetAlgoLined
 \KwIn{$\mathcal{D}$, $\epsilon$,$L$,$K$}
 \KwOut{$\theta_q$, $\theta_s$}
 Randomly initialize parameters $\theta_q,\theta_s \leftarrow \text{Uniform}(-0.1,0.1)$;\\
{\bf $\#$Pretraining the S-Network.}\\
\For{$j=1:K$}{
Sample random mini-batches of $(s_t,i_t,r_t,s_{t+1})$ from $\mathcal{D}$;\\
Set $f_t,d_t,v^r,l_t$ according to $s_t,r_t,s_{t+1}$;\\ 
Update $\theta_s$ via mini-batch SGD w.r.t. the loss in Equation~(\ref{equ:loss});\\
}
{\bf $\#$ Iterative training of S-Network and Q-Network.};\\
\Repeat{convergence}{
    \For{$j=1:N$}{
    {\bf $\#$ Sampling training samples from logged data.}\\
    Sampling $(s,i,r,s')$ from $\mathcal{D}$, and storing in buffer $\mathcal{M}$;\\
    {\bf $\#$ Sampling training samples by interacting with the S-Network.}\\
    $l = \text{False}$;\\
    sample a initial user $u$ from user set;\\
    initial $s = \{u\}$;\\
    \While{$l$ is False}{
    sample a recommendation $i$ \emph{w.r.t} $\epsilon$-greedy Q-value;\\
    execute $i$;\\
    S-Network responds with $f$, $d$, $l$,$v^r$;\\
    set $r$ according to $f$, $d$, $l$,$v^r$;\\
    set $s'= s \oplus \{i,r,d\}$;\\
    store $(s,i,r,s')$ in buffer $\mathcal{M}$;\\
    update $s\leftarrow s'$;\\
    }
    {\bf $\#$ Updating the Q-Network.}\\
    \For{$j=1:L$}{
    Sample random mini-batches of training $(s_t,i_t,r_t,s_{t+1})$ from $\mathcal{M}$;\\
    Update $\theta_{q}$ via mini-batch SGD w.r.t. Equation~(\ref{equ:qlearning});\\
    }
    {\bf $\#$ Updating the S-Network.}\\
    \For{$j=1:K$}{
    Sample mini-batches of $(s_t,i_t,r_t,s_{t+1})$ from $\mathcal{M}$;\\
    Set $f$, $d$, $l$,$v^r$ according to $r_t$, $s_{t+1}$;\\
    Update $\theta_s$ via mini-batch SGD \emph{w.r.t.} the loss in Equation~(\ref{equ:loss});\\
    }
}}
\caption{Offline training of FeedRec.}
\label{alg:training}
\end{algorithm}
}

\subsection{Simulator Learning}
In this process, $S(s_t,i_t;\theta_s)$ is refined via mini-batch SGD using logged data in the $\mathcal{D}$. As the logged data is collected via a logging policy $\pi_b$, directly using such logged data to build the simulator will cause the selection base. To debias the effects of loggind policy $\pi_b$~\cite{schnabel2016recommendations}, an importance weighted loss is minimized as follow:
{\small
\begin{eqnarray}\label{equ:loss}
    \ell(\theta_s) &=& \sum_{t=0}^{T-1}\gamma^t \frac{1}{n}\sum_{k=1}^{n}\{\omega_{0:t},c\}\delta_t(\theta_s)\\
    \nonumber \delta_t(\theta_s) &=&  \lambda_f\cdot \Psi(\bm{f}_t,\hat{\bm{f}}_t)+\lambda_d \cdot (d_t-\hat{d}_t)^2 +\\
    \nonumber     &&\lambda_l\cdot \Psi(l_t,\hat{l}_t) + \lambda_v\cdot (v^r-\hat{v}^r)^2,
\end{eqnarray}}
where $n$ is the total number of trajectories in the logged data. $\omega_{0:t} = \prod_{k=0}^t\frac{\pi(i_k|s_k)}{\pi_b(i_k|s_k)}$ is the importance ratio to reduce the disparity between $\pi$ (the policy derived from Q-Network, \eg $\epsilon$-greedy) and $\pi_b$, $\Psi(\cdot,\cdot)$ is the cross-entropy loss function, and $c$ is a hyper-parameter to avoid too large importance ratio. $\delta_t(\theta_s)$ is a multi-task loss function that combines two classification loss and two regression loss, $\lambda_\ast$ is the hyper-parameter controling the importance of different task.  

As $\pi$ derived from Q-Network is constantly changed with the update of $\theta_q$, to keep adaptive to the intermediate policies, the S-Network also keep updated in accordance with $\pi$ to obtain the customized optimal accuracy. Finally, we implement an interactive training procedure, as shown in Algorithm \ref{alg:training}, where we specify the order in which they occur within each iteration. 

\section{simulation study}\label{simulation}
We demonstrate the ability of FeedRec to fit the user's interests by directly optimizing user engagement metrics through simulation.
We use the synthetic datasets so that we know the ``ground truth'' mechanism to maximize user engagement, and we can easily check whether the algorithm can indeed learn the optimal policy to maximize delayed user engagement.

\subsection{Setting}
Formally, we generate $M$ users and $N$ items, each of which is associated with a $d$-dimensional topic parameter vector $\bm{\vartheta}\in \mathbb{R}^d$.
For $M$ users ($\mathcal{U}=\{\bm{\vartheta}_u^{(1)},\dots,\bm{\vartheta}_u^{(M)}\}$) and $N$ items ($\mathcal{I}=\{\bm{\vartheta}_i^{(1)},\dots,\bm{\vartheta}_i^{(N)}\}$), the topic vectors are initialized as
{\small
\begin{eqnarray}\label{eq:user_vector_set}
\bm\vartheta=\frac{\tilde{\bm\vartheta}}{||\tilde{\bm\vartheta}||}, \mathrm{where}\ \tilde{\bm\vartheta}=\left\{
\begin{aligned}
&\tilde{\vartheta}_{k}=1-\kappa , & \mathrm{the\ primary\ topic}\ k,\\
&\tilde{\vartheta}_{k'}\sim U(0,{\kappa} ), & \ k'\neq k,
\end{aligned}
\right.
\end{eqnarray}}
where $\kappa$ controls how much attention would be paid on non-primary topics.
Specifically, we set the dimension of user vectors and item vectors to 10.
Once the item vector $\bm{\vartheta}_i$ is initialized, it will keep the same for the simulation.
At each time step $t$, the agent feeds one item $\bm{\vartheta}_i$ from $\mathcal{I}$ to one user $\bm{\vartheta}_u$.
The user checks the feed and gives feedback, \eg click/skip, leave/stay (depth metric), and revisit (return time metric), based on the ``satisfaction''.
Specifically, the probability of click is determined by the cosine similarity as $p(\text{click}|\bm{\vartheta}_u,\bm{\vartheta}_i) = \frac{\bm{\vartheta}_i^\top \bm{\vartheta}_u}{\|\bm{\vartheta}_i\|\|\bm{\vartheta}_u\|}$.
For leave/stay or revisit, these feedback are related to all the feeds.
In the simulation, we assume these feedback are determined by the mean entropy of recommendation list because many existing works\cite{Adomavicius:AggregateDIV,Ashkan:GreedyDIV,Cheng:AccurateDiversity} assume the diversity is able to improve the user's satisfactory on the recommendation results.
Also, diversity is also delayed metrics~\cite{Xia:MDPsearch,zou2019reinforcement}, which can verify whether FeedRec could optimize the delayed metrics or not.

\subsection{Simulation Results}

Some existing works \cite{Adomavicius:AggregateDIV,Ashkan:GreedyDIV,Cheng:AccurateDiversity} assume the diversity is able to improve the user's satisfactory on the recommendation results.
Actually, it is an indirect method to optimize user engagement, and diversity here play an instrumental role to achieve this goal.
We now verify that the proposed FeedRec framework has the ability to directly optimize user engagement through different forms of diversity.
To generate the simulation data, we follow the popular diversity assumption\cite{Adomavicius:AggregateDIV,Ashkan:GreedyDIV,Cheng:AccurateDiversity}. 
These works tried to enhance diversity to achieve better user engagement.
However, it is unclear that to what extent the diversity will lead to the best user engagement.
Therefore, the pursuit of diversity may not lead to the improvement of user satisfaction.
 
\begin{figure}[!t]
\includegraphics[width=8.0cm]{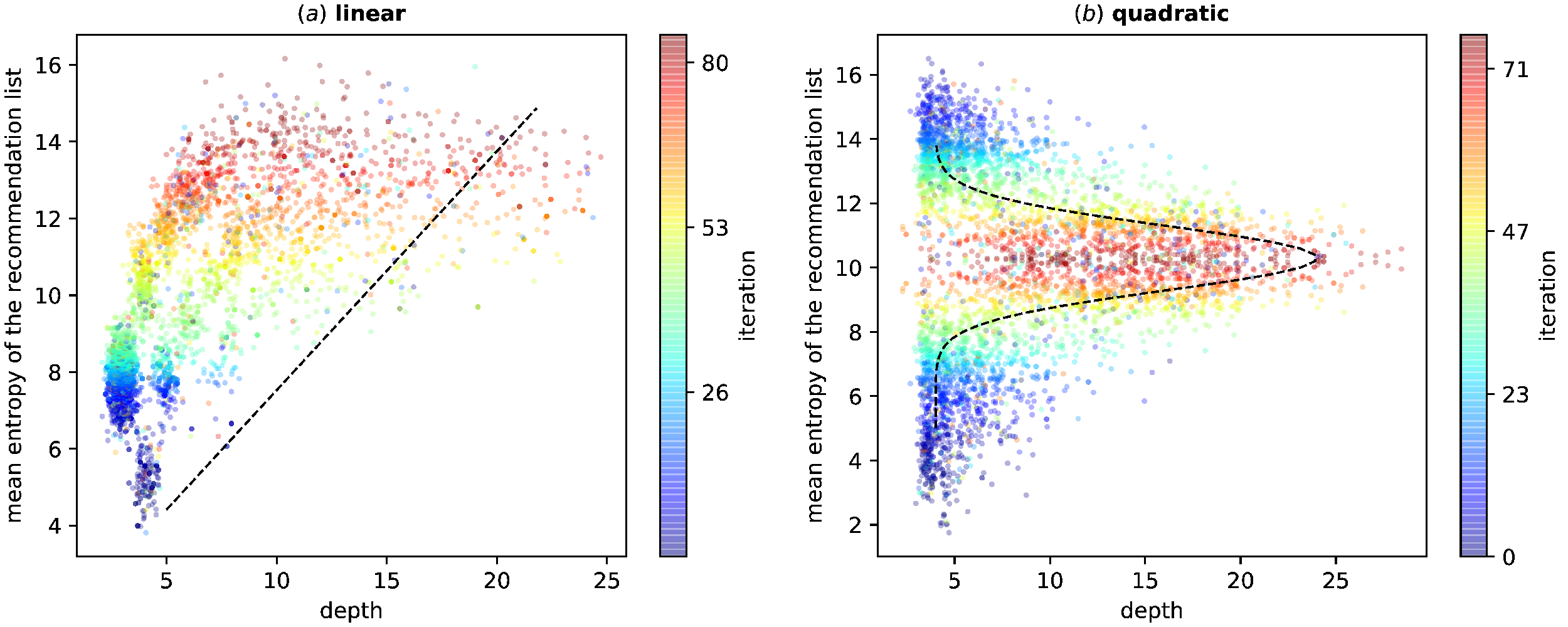}
\caption{\small Different distributions of user's interests and browsing depth. The dashed line represents the distribution of scrolling down and entropy (linear in (a) and quadratic in (b)). The color bar shows the interaction iteration in training phrase, from blue to red. The average browsing depth over all users are shown as dots.}
\label{fig:simulation_depth}
\end{figure}

\begin{figure}[!t]
\includegraphics[width=8.0cm]{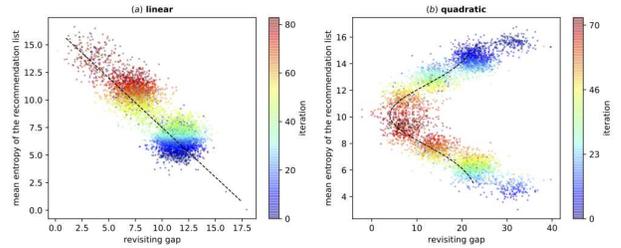}
\caption{\small Different distributions of user's interests and interval days between two visits. The dashed line represents the distribution of return time and entropy (linear in (a) and quadratic in (b)). The color bar shows the interaction iteration in training phrase, from blue to red. The average return time over all users are shown as dots.}
\label{fig:simulation_returning}
\end{figure}

We assume that there are two types of relationship between user engagement and diversity of recommendation list. 

1) {\bf Linear style}.
In the linear relationship, higher entropy brings more satisfaction, that is, higher entropy attracts user to browse more items and use the system more often.
The probability of user staying with the system after checking the fed items is set as:
{\small
   \begin{eqnarray*}
    p(stay|\bm{\vartheta}_{1},\dots,\bm{\vartheta}_{t}) &=& a \mathfrak{E}(\bm{\vartheta}_{1},\dots,\bm{\vartheta}_{t}) + b, a>0 \\
       \mathfrak{E}(\bm{\vartheta}_{1},\dots,\bm{\vartheta}_{t}) &=& \frac{1}{t\times(t-1)}\sum_{ \substack{m,n\in\{1,\dots,t\} \\ m\neq n}}\bm{\vartheta}_{m}\log\frac{\bm{\vartheta}_{m}}{\bm{\vartheta}_{n}}
  \end{eqnarray*}}
where $\{\bm{\vartheta}_{1},\dots,\bm{\vartheta}_{t}\}$ is the list of recommended items, $\mathfrak{E}(\bm{\vartheta}_{1},\dots,\bm{\vartheta}_{t})$ is the mean entropy of the items. $a$ and $b$ are used to scale into range (0,1). The interval days of two visit is set as:
{\small
\begin{eqnarray*}
    v^{r} = V - d*\mathfrak{E}(\bm{\vartheta}_{i,1},\dots,\bm{\vartheta}_{i,t}), V> 0, d>0,
\end{eqnarray*}}
where $V$ and $d$ are constants to make $v^r$ positive.

2) {\bf Quadratic style}.
In the quadratic relationship, moderate entropy makes the best satisfaction.
The probability of user staying with the system after checking the fed items is set as: 
{\small
\begin{eqnarray*}
    p(stay|\bm{\vartheta}_{1},\dots,\bm{\vartheta}_{t}) = \exp\{-\frac{(\mathfrak{E}(\bm{\vartheta}_{1},\dots,\bm{\vartheta}_{t})-\mu)^2}{\sigma}\},
\end{eqnarray*}}
where $\mu$ and $\sigma$ are constants.
Similarly, the interval days of two visit is set as:
{\small
\begin{eqnarray*}
    v^r = V (1- \exp\{-\frac{(\mathfrak{E}(\bm{\vartheta}_{i,1},\dots,\bm{\vartheta}_{i,t})-\mu)^2}{\sigma}\}) , V> 0.
\end{eqnarray*}}
Following the above process of interaction between ``user'' and system agent, we generate 1,000 users, 5,000 items, and 2M episodes for training.

We here report the average browsing depth and return time w.r.t. different relationship---linear or quadratic---of each training step, where the blue points are at earlier training steps and the red point is at the later training steps.
From the results shown in Figure \ref{fig:simulation_depth} and Figure \ref{fig:simulation_returning}, we can see that no matter what diversity assumptions are, FeedRec is able to converge to the best diversity by directly optimizing delay metrics.
In (a) of Figure \ref{fig:simulation_depth} (and also Figure \ref{fig:simulation_returning}), FeedRec discloses that the distribution of entropy of recommendation list and browsing depth (and also return time) is linear.
As the number of rounds of interaction increases, the user's satisfaction gradually increases, therefore more items are browsed (and internal time between two visits is shorter).
In (b), user engagement is highest in a certain entropy, higher or lower entropy can cause user's dissatisfaction.
Therefore, moderate entropy of recommendation list will attract the user to browse more items and use the recommender system more often.
The results indicate that FeedRec has the ability to fit different types of distribution between user engagement and the entropy of the recommendation list.

\section{Experiments on Real-world E-commerce Dataset}\label{experiments}

\subsection{Dataset}
We collected 17 days users' accessing logs in July 2018 from an e-commerce platform.
Each accessing log contains: timestamp, user id $u$, user's profile ($\bm{u}_p \in\mathbb{R}^{20}$), recommended item's id $i_t$, behavior policy's ranking score for the item $\pi_b(i_t|s_t)$, and user's feedback $f_t$, dwell time $d_t$.
Due to the sparsity of the dataset, we initialized the items' embedding ($\bm{i}\in\mathbb{R}^{20}$) with pretrained vectors, which is learned through modeling users' clicking streams with skip-gram~\cite{Mikolov:word2vec}.
The user's embedding $\bm{u}$ is initialized with user's profile $\bm{u}_p$.
The returning gap was computed as the interval days of the consecutive user visits.
Table \ref{tab:data_info} shows the statistics of the dataset.

\begin{table}[!t]
 	\caption{\small Summary statistics of dataset.}
 	\begin{center}
 	\small
 		\begin{tabular}
 			{l|l}
 			\Xhline{1.2pt}
 			Statistics &  Numerical Value \\
 			\Xhline{1.2pt}
 			The number of trajectories & 633,345\\
 			The number of items & 456,805 \\
 			The number of users & 471,822 \\
 			Average/max/min clicks & 2.04/99/0\\
 			Average/max/min dwell time(minutes) & 2.4/5.3/0.5 \\
 			Average/max/min browsing depth & 13.34/149/1 \\
 			Average/max/min return time (days) & 5.18/17/0\\
 			\Xhline{1.2pt}
 		\end{tabular}
 	\end{center}
 	\label{tab:data_info}
\end{table}

\subsection{Evaluation Setting}
\paragraph{\bf off-line A/B testing} To perform evaluation of RL methods on ground-truth, a straightforward way is to evaluate the learned policy through online A/B test, which, however, could be prohibitively expensive and may hurt user experiences. Suggested by~\cite{Gilotte:offlineAB,farajtabar:more}, a specific off-policy estimator NCIS~\cite{Swaminathan:NIS} is employed to evaluate the performance of different recommender agents.
The step-wise variant of NCIS is defined as
\begin{eqnarray}\label{equ:evaluation}
    \hat{R}^{\pi}_{step-NCIS} &=& \sum_{\xi_k \in \mathcal{T}} \sum_{t=0}^{T-1} \frac{\bar{\rho}^{i}_{0:t}r_t^{k}}{\sum_{j=1}^{K}\bar{\rho}_{0:t}^{j}}\\
\nonumber \bar{\rho}_{t_1:t_2} &=& \min \{c,\prod_{t=t_1}^{t_2}\frac{\pi(a_t|s_t)}{\pi_b(a_t|s_t)}\}, 
\end{eqnarray}
where $\bar{\rho}_{t_1:t_2}$ is the max capping of importance ratio, $\mathcal{T}=\{\xi_k\}$ is the set of trajectory $\xi_k$ for evaluation, $K$ is the total testing trajectory. The numerator of Equation (\ref{equ:evaluation}) is the capped importance weighted reward and the denominator is the normalized factor. Setting the $r_t$ with different metrics, we can evaluate the policy from different perspective. To make the experimental results trustful and solid, we use the first 15 days logging as training samples, the last 2 days as testing data, the test data is kept isolated. The training samples are used for policy learning. The testing data are used for policy evaluation. To ensure small variance and control the bias, we set $c$ as 5 in experiment.

\paragraph{\bf The metrics}
Setting the reward in Equation (\ref{equ:evaluation}) with different user engagement metrics, we could estimate a comprehensive set of evaluation metrics. Formally, these metrics are defined as follows,
\begin{itemize}[leftmargin=*]
	\item Average Clicks per Session: the average cumulative number of clicks over a user visit.
	\item Average Depth per Session: the average browsing depth that the users interact with the recommender agent.
	\item Average Return Time: the average revisiting days between a user's consecutive visits up till a particular time point.
\end{itemize}
\noindent
\paragraph{\bf The baselines}
We compare our model with state-of-the-art baselines, including both supervised learning based methods and reinforcement learning based methods.
\begin{itemize}[leftmargin=*]
    \item \textbf{FM}: Factorization Machines ~\cite{Rendle:FM} is a strong factoring model, which can be easily implemented with factorization machine library (libFM)\footnote{http://www.libfm.org}.
    \item \textbf{NCF}: Neural network-based Collaborative Filtering~\cite{He:Nueral} replaces the inner product in factoring model with a neural architecture to support arbitrary function from data.
    \item \textbf{GRU4Rec}: This is a representative approach that utilizes RNN to learn the dynamic representation of customers and items in recommender systems~\cite{Hidasi:sessionRNN}.
    \item \textbf{NARM}: This is a state-of-the-art approach in personalized trajectory-based recommendation with RNN models~\cite{Li:NARM}. It uses the attention mechanism to determine the relatedness of the past purchases in the trajectory for the next purchase.  
    \item \textbf{DQN}: Deep Q-Networks~\cite{Mnih:DQN} combined Q-learning with Deep Neural Networks. We use the same function approximation as FeedRec and train the neural network with naive Q-learning using the logged dataset. 
    \item \textbf{DEERs}: DEERs~\cite{zhao2018Pairwise}  is a DQN based approach for maximizing users' clicks with pairwise training, which considers user's negative behaviors.
    \item \textbf{DDPG-KNN}: Deep Deterministic Policy Gradient with KNN~\cite{Dulac:RL_large_action} is a discrete version of DDPG for dealing with large action space, which has been deployed for Pagewise recommendation in~\cite{zhao:deepDPG}.
    \item \textbf{FeedRec}: To verify the effect of different components, experiments are conducted on the degenerated models as follow: 1)~ \textbf{S-Network} is purely based on our proposed S-Network, which makes recommendations based on the ranking of next possible clicking item. 2)~\textbf{FeedRec(C)}, \textbf{FeedRec(D)}, \textbf{FeedRec(R)} and \textbf{FeedRec(All)} are our proposed methods with different metrics as reward. Specifically, they use the clicks, depth, return time and the weighted sum of instant and delayed metrics as the reward function respectively. 
\end{itemize}

\paragraph{\bf Experimental Setting}
The weight $\bm{\omega}$ for different metrics is set to $[1.0,0.005,0.005]^{\top}$. The hidden units for LSTM is set as 50 for both Q-Network and S-Network. All the baseline models share the same layer and hidden nodes configuration for the neural networks. The buffer size for Q-Learning is set 10,000, the batch size is set to 256. $\epsilon$-greedy is always applied for exploration in learning, but discounted with increasing training epoch. The value $c$ for clipping importance sampling is set 5. We set the discount factor $\gamma = 0.9$. The networks are trained with SGD with learning rate of 0.005. We used Tensorflow to implement the pipelines and trained networks with a Nvidia GTX 1080 ti GPU cards. All the experiments are obtained by an average of 5 repeat runs.

\subsection{Experimental Results}
\begin{table}
\small
\tabcolsep 0.03in
\centering
\caption{\small Performance comparison of different agents on JD dataset.}
\label{tab:results}
\begin{tabular}{l| c c c} \ChangeRT{0.8pt}
Agents & \tabincell{c}{Average Clicks\\ per Session} & \tabincell{c}{Average Depth\\ per Session} & \tabincell{c}{Average\\ Return Time} \\ \hline \hline
FM                & 1.9829 & 11.2977 & 16.5349  \\
NCF               & 1.9425 & 11.1973 & 18.2746  \\ \hline \hline
GRU4Rec           & 2.1154 & 13.8060 & 14.0268 \\
NARM              & 2.3030 & 15.3913 & 11.0332 \\ \hline \hline
DQN               & 1.8211 & 15.2508 & 6.2307 \\
DEER              & 2.2773 & 18.0602 & 5.7363 \\
DDPG-KNN(k=1)     & 0.6659 & 9.8127 & 15.4012 \\
DDPG-KNN(k=0.1N)  & 2.5569 & 16.0936 & 7.3918 \\
DDPG-KNN(k=N)     & 2.5090 & 14.6689 & 14.1648 \\\hline \hline
S-Network         & 2.5124 & 16.1745 & 10.1846 \\ 
FeedRec(C)        & 2.6194 & 18.1204 & 6.9640 \\
FeedRec(D)        & 2.8217 & 21.8328 & 4.8756 \\
FeedRec(R)        & 3.7194 & 23.4582 & 3.9280 \\
FeedRec(All)      & $\textbf{4.0321}^\ast$ & $\textbf{25.5652}^\ast$ & $\textbf{3.9010}^\ast$ \\
\ChangeRT{0.8pt}
\end{tabular}
\begin{tablenotes}
\centering
\item $``\ast"$ indicates the statistically significant improvements (\ie two-sided $t$-test with  $p<0.01$) over the best baseline.
\end{tablenotes}
\end{table}

\paragraph{\bf Comparison against baselines}
We compared FeedRec with state-of-the-art methods.
The results of all methods over the real-world dataset in terms of three metrics are shown in Table~\ref{tab:results}.
From the results, we can see that FeedRec outperformed all of the baseline methods on all metrics.
We conducted significance testing (t-test) on the improvements of our approaches over all baselines.
$``*"$ denotes strong significant divergence with \text{p-value}<0.01.
The results indicate that the improvements are significant, in terms of all of the evaluation measures.
\begin{figure}[!t]
\includegraphics[width=8.5cm]{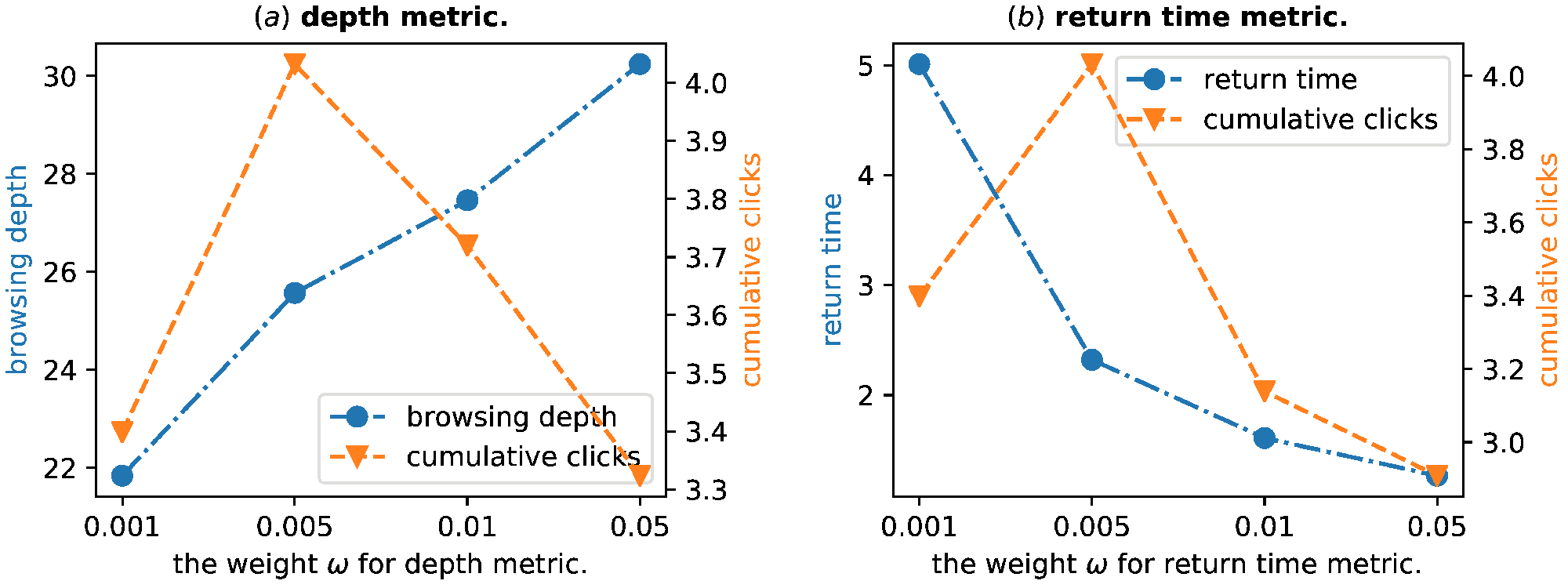}
\caption{\small The influence of $\omega$ on performance.}
\label{fig:reward_influence}
\end{figure}

\paragraph{\bf The influence of weight $\bm{\omega}$.}
The weight $\bm{\omega}$ controls the relative importance of different user engagement metrics in reward function.
We examined the effects of the weights $\omega$.
(a) and (b) of Figure \ref{fig:reward_influence} shows the parameter sensitivity of $\omega$ \emph{w.r.t.} depth metric and return time metric respectively.
In Figure \ref{fig:reward_influence}, \emph{w.r.t.} the increase of the weight of $\omega$ for depth and return time metrics, the user browses more items and revisit the application more often (the blue line). Meanwhile, in both (a) and (b), the model achieves best results in the cumulative clicks metric (the orange line) when the $\omega$ is set to 0.005. Too much weight on these metrics will overwhelm the importance of clicks on the rewards, which indicates that moderate value of weights on depth and return time can indeed improve the performance on cumulative clicks.



\begin{figure}[!t]
\includegraphics[width=9.0cm]{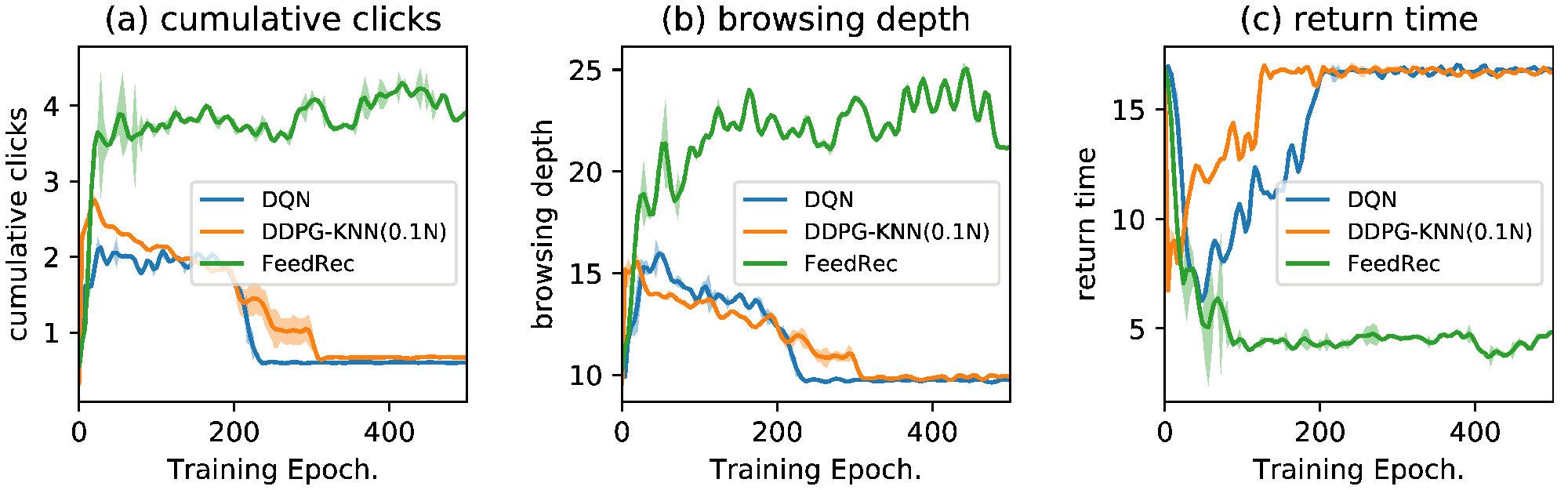}
\caption{\small Comparison between FeedRec and baselines under offline learning.}
\label{fig:q_learning_influence}
\end{figure}
  
\paragraph{\bf The effect of S-Network}
The notorious {\it deadly triad} problem causes the danger of instability and divergence of most off-policy learning methods, even the robust Q-Learning.
To examine the advantage of our proposed interactive training framework, we compared our proposed model FeedRec with DQN, DDPG-KNN under the same configuration.
In Figure \ref{fig:q_learning_influence}, we show different metrics vs the training iteration.
We find that DQN, DDPG-KNN achieves a performance peak around 40 iterations and the performances are degraded rapidly with increasing iterations (the orange line and blue line).
On the contrary, FeedRec achieves better performances on these three metrics and the performances are stable at the highest value (the green line).
These observations indicate that FeedRec is stable and suitable through avoiding the {\it deadly triad} problem for off-policy learning of recommendation policies.


\paragraph{\bf The relationship between user engagement and diversity}

\begin{figure}[!t]
\includegraphics[width=8cm]{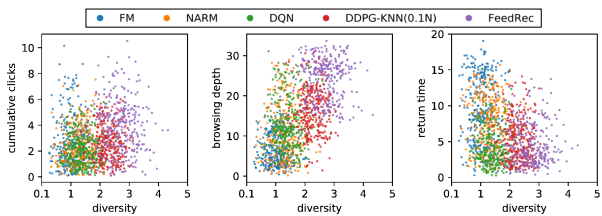}
\caption{\small The relationship between user engagement and diversity.}
\label{fig:engagement_diversity}
\end{figure}

Some existing works~\cite{Adomavicius:AggregateDIV,Ashkan:GreedyDIV,Cheng:AccurateDiversity} assume user engagement and diversity are related and intent to increase user engagement by increasing diversity. Actually, it is an indirect method to optimize the user engagement, and the assumption has not been verified. Here, we conducted experiments to see whether FeedRec, which direct optimize user engagement, has the ability to improve the recommendation diversity. For each policy, we sample $300$ state-action pairs with importance ratio $\bar{\rho}>0.01$ (the larger value of $\bar{\rho}$ in Equation (\ref{equ:evaluation}) implies that the policy more favors such actions) and plot these state-action pairs, which are shown in Figure \ref{fig:engagement_diversity}. The horizontal axis indicates the diversity between recommendation items, and the vertical axis indicates different types of user engagement~(\eg browsing depth, return time). We can see that the FeedRec policy, learned by directly optimizing user engagement, favors for recommending more diverse items. The results verifies that optimization of user satisfaction can increase the recommendation diversity and enhancing diversity is also a means of improving user satisfaction.

\section{Conclusion}\label{Conclusion}
It is critical to optimize long-term user engagement in the recommender system, especially in feed streaming setting.
Though RL naturally fits the problem of maximizing the long-term rewards, there exist several challenges for applying RL in optimizing long-term user engagement: difficult to model the omnifarious user feedbacks (\eg clicks, dwell time, revisit, \etc) and effective off-policy learning in recommender system.
To address these issues, in this work, we introduce a RL-based framework --- FeedRec to optimize the long-term user engagement.
First, FeedRec leverage hierarchical RNNs to model complex user behaviors, refer to as Q-Network. Then to avoid the instability of convergence in policy learning, an S-Network is designed to simulate the environment and assist the Q-Network.
Extensive experiments on both synthetic datasets and real-world e-commerce dataset have demonstrated effectiveness of FeedRec for feed streaming recommendation.
\bibliographystyle{ACM-Reference-Format}
\bibliography{main.bib} 

%

\end{document}